\documentclass[]{interact}

\usepackage{epstopdf}
\usepackage{subfigure}

\usepackage[numbers,sort&compress]{natbib}
\bibpunct[, ]{[}{]}{,}{n}{,}{,}

\theoremstyle{plain}

\theoremstyle{definition}

\theoremstyle{remark}

\begin{document}

\title{Nonlinear transport theory in the metal with tunnel barrier}

\author{
\name{E.~E. Zubov\textsuperscript{a,b}\thanks{CONTACT E.~E. Zubov. Email: eezubov@ukr.net}}
\affil{\textsuperscript{a}G.V. Kurdyumov Institute for Metal Physics, NASU, 03680 Kyiv, Acad. Vernadsky Boulevard 36, Ukraine; \textsuperscript{b}Vasyl' Stus Donetsk National University, 021021, Vinnytsia, Ukraine}
}

\maketitle

\begin{abstract}
Within the framework of the scattering  matrix  formalism  the nonlinear Kubo theory for electron transport in the metal with a tunnel barrier has been considered. A general expression for the mean electrical current was obtained. It significantly simplifies the calculation of nonlinear contributions to the conductivity of various hybrid structures. In the model of the tunnel Hamiltonian all linear and nonlinear contributions to a mean electrical current are evaluated. The  linear approximation  agrees with results of other theories.  For  effective barrier transmission $\tilde T = 1/3$  the ballistic transport is  realized with  a value of the Landauer conductivity equal to $G = \frac{{2e^2 }}{h}$.
\end{abstract}


\section{Introduction}

At present the method for solving a problem of electron transport in the hybrid structures is    one of the key tasks  from the theoretical as well as the practical point of view. On the one hand, a  derivation  of  the theory in the framework of modern diagrammatic methods based on the scattering matrix formalism makes it possible to better understand the nature of electronic dynamics in  nonequilibrium processes and also to study nonlinear effects in condensed matter. On the other hand, the design of the electron devices demands the sufficiently detailed information on processes of the electron scattering on contacts, barriers and impurities to be caused by technology of  industry. Also, in the miniaturization of working medium, the problem of accounting for quantum properties in nanocontacts and nanowires somehow arises.

It should be noted that the researchers have accumulated a considerable experience in solving these problems \cite{A1,A2,A3,A4,A5} for both normal metals and superconductors. Nevertheless, there are still a number of unresolved questions related to the effects of nonlinear contributions to the current-voltage characteristics of various solid-state structures.
One of the problems is related to the fact that in  the  calculation of  a  mean current, it is necessary to take into account the time-dependent evolution of a wave function and a dependence of current operator on the real time $t$. Unfortunately, the developed method of Matsubara \cite{A6} for evaluating a mean value of the operators at \textit{T}$\neq 0$ implies a transition to the imaginary time. Indeed, in this case the density matrix is used which satisfies the Bloch equation. As a result, a formula for calculating the \textit{S}-matrix contains integrals over the imaginary time $\tau$ = $\textit{it}$. This discrepancy raises a problem of the passage to the limiting case of zero temperatures.
Therefore, we will study this question in more detail, which will allow us to generalize the Kubo theory for transport in the nonlinear limit using the \textit{S}-matrix. In \cite{A7}, in studying the statistical mechanics of irreversible processes R. Kubo presented a general expression for evolution of the density matrix in the quantum-mechanical case, which was expressed in terms of commutators of the perturbation operators in the interaction representation with unperturbed density matrix. However, the obtained formula is technically very difficult to use because of the presence of complicated commutators. Therefore, in the majority of cases \cite{A8,A9} only a linear response   is considered that does not give a complete picture of the influence of nonlinear contributions to the transport.

Taking into account the above, the structure of the work is as follows. Section 2 develops the Kubo theory for the chronological products of perturbation operators. This allows us to do away with complex commutators of the time-dependent operators  and also to obtain a simplified Kubo formula for the mean current. Section 3 discusses the specific application of the resulting formula for the mean current in a hybrid metal structure with a tunnel barrier in the linear approximation. The results obtained are consistent with the already known data. In Section 4, the theory is further generalized to the contribution to the mean current in any order of time perturbation theory. A quantum-statistical correction is obtained to the already known Landauer formula for conductance, It is also shown that the tunneling Hamiltonian correctly describes the transport properties of  hybrid structures.

\section{ Scattering matrix formalism in the nonlinear Kubo theory for electron transport}

In this section, we consider the effect of the barrier and the applied voltage on the electron transport in a metal at the nonzero  temperature \textit{T}. In the framework of perturbation theory, the evolution of the wave function in the interaction representation is determined by the perturbation only. It simplifies the calculations considerably. Taking into account the above, we start with the Liouville equation for the evolution of the density matrix in real time ( $\hbar=1$):
\begin{eqnarray}
i\frac{{\partial \rho (t)}}{{\partial t}} = \left[ {H_T (t),\rho (t)} \right]
\,,
\label{energy01}
\end{eqnarray}
where $\rho (t) = e^{ - i\hat H_0 t} \rho _I (t)e^{i\hat H_0 t}$, $\rho _I (t)$ and $H_T (t)$ are the density matrix $\rho (t)$  and perturbation Hamiltonian $H_T$    in the interaction representation. Also, it is easy to show that 
\begin{eqnarray}
\rho _I (t) = \tilde U(t)\rho _0 \tilde U^ +  (t),
\label{energy02}
\end{eqnarray}
where the operator $\tilde U(t) = e^{i\hat H_0 t} e^{ - i\hat Ht}$. The operator $\rho _0  = \frac{{e^{ - \beta \hat H_0 } }}{{Tr(e^{ - \beta \hat H_0 } )}}$ is a standard expression for unperturbed density matrix at ${\textit{t} =  - }\infty$ and ${\rm 1/}\beta {\textit{= T}}$ is the temperature. Here, it is supposed an infinite slow  switching on a  perturbation $H_T$ to the unperturbed part $\hat H_0$ up to full Hamiltonian $\hat H$. The Schr${\rm \ddot o}$dinger evolution operator $U(t) = e^{ - i\hat H_0 t} \tilde U(t)$ of the wave function, where $\tilde U(t){\rm }$ is expressed in terms of scattering matrix ${\textit{S(t,}-\infty)}$ by the next manner
\begin{eqnarray}
\label{energy03}
\tilde U(t) = S(t, - \infty ) = T_t \exp \left[ { - i\int\limits_{ - \infty }^t {H_T (t)dt} } \right]{\rm  },
\end{eqnarray}
Here, \textit{T}$_{\textit{t}}$ is the time-ordered operator. Apparently, that the Hermitian conjugate operator $U^ +  (t) = \tilde U^ +  (t)e^{i\hat H_0 t}$. Then the mean current is determined by expression
\begin{eqnarray}
\label {energy04}
\begin{array}{l}
  < I >  = Tr(\rho I) = \frac{1}{{\sum\limits_n {\left\langle n \right|U^ +  (t)} \rho _0 U(t)\left| n \right\rangle }}\sum\limits_n {\left\langle n \right|U^ +  (t)IU(t)} U^ +  (t)\rho _0 U(t)\left| n \right\rangle  =  \\ 
 \quad \quad \frac{1}{{Tr(\tilde U(t)\rho _0 \tilde U^ +  (t))}}Tr\left( {e^{i\hat H_0 t} Ie^{ - i\hat H_0 t} \tilde U(t)\rho _0 \tilde U^ +  (t)} \right), \\ 
 \end{array}
\end{eqnarray}
where  \textit{I}  is the current operator and the symbol $Tr = \sum\limits_n {\left\langle n \right|} ...\left| n \right\rangle$   denotes trace with summation over complete set of states. By using both the cyclic property of the trace and the definition of the current operator in the interaction representation, we write the expression for a mean current in the form:
\begin{eqnarray}
\label{energy05}
 < I >  = \frac{1}{{Sp(\rho _I (t))}}Sp(I(t)\rho _I (t))
\end{eqnarray}
It is easy to present the expression ~(\ref{energy02}) for $\rho _I (t)$   as a series using the Baker-Campbell-Hausdorff  formula \cite{A10}: 
\begin{eqnarray}
\label{energy06}
 e^A Be^{ - A}  = \sum\limits_{j = 0}^\infty  {\frac{1}{{j!}}} \left[ {A,B} \right]_{(j)}, 
\end{eqnarray}
where the symbol $\left[ {A,B} \right]_{(j)}$ denotes the \textit{k}-order commutator determined in accordance with recurrence relation
\begin{eqnarray}
\label{energy07}
\left[ {A,B} \right]_{(j + 1)}  = \left[ {A,\left[ {A,B} \right]_{(j)} } \right]
\end{eqnarray}
At that $\left[ {A,B} \right]_{(0)}  = B$ and $\left[ {A,B} \right]_{(1)}  = \left[ {A,B} \right] = AB - BA$ is the ordinary commutator. Assuming that the perturbation ${H_T}$    is a Hermitian operator, we write the final expression for $\rho _I (t)$ in any order of perturbation theory:
\begin{eqnarray}
\label{energy08}
\begin{array}{l}
\rho _I (t) = \\ \sum\limits_{j = 0}^\infty  {\frac{{( - i)^j }}{{j!}}} \int\limits_{ - \infty }^t {dt_j dt_{j - 1} ...dt_1 T_t \left\{ {\left[ {H_T (t_j ),\left[ {H_T (t_{j - 1} ),\left[ {...} \right[H_T (t_2 ),\left[ {H_T (t_1 ),\rho _0 } \right]} \right]...} \right]} \right\}}
 \end{array} 
\end{eqnarray}
In the first order in   from Eq.~(\ref{energy02}) it is easy to obtain the linear response in the framework of the Kubo theory for charge transport. A generalization to higher orders of perturbation theory causes certain difficulties due to the presence of a large number of commutators under the signs of integrals. It considerably complicates their calculation. Therefore, for convenience, we transform the integrand ~(\ref{energy08}). It is obvious that for any \textit{t}${_j}$    the identity holds
\begin{eqnarray}
\label{energy09}
\left[ {H_T (t_j ),\rho _0 } \right] = H_T (t_j )\rho _0  - \rho _0 H_T (t_j )\rho _0^{ - 1} \rho _0  = A(t_j )\rho _0,
\end{eqnarray}
where
\begin{eqnarray}
\label{energy10}
A(t_j ) = H_T (t_j ) - \rho _0 H_T (t_j )\rho _0^{ - 1} 
\end{eqnarray}

Let consider the commutator in the second  order of the expansion ~(\ref{energy08}):
\begin{eqnarray}
\label{energy11}
\begin{array}{l}
{\rm   }\left[ {H_T (t_2 )\left[ {H_T (t_1 ),\rho _0 } \right]} \right] = \\ A(t_1 )\left[ {H_T (t_2 ),\rho _0 } \right] + \left[ {H_T (t_2 ),A(t_1 )} \right]\rho _0  = A(t_1 )A(t_2 )\rho _0  + \left[ {H_T (t_2 ),A(t_1 )} \right]\rho _0 
 \end{array} 
\end{eqnarray}
The last term in Eq.~(\ref{energy11}) under the signs of the integrals and \textit{T}${_t}$   equals identically zero, since the operator ${H_T}(t_2)$  is a Bose type.  Thus,   the sign of the commutator term with minus sign  after  time ordering is not changed. Obviously, the commutators of higher orders are simplified in a similar way. Then Eq.~(\ref{energy08}) can be written in the form
\begin{eqnarray}
\label{energy12}
\rho _I (t) = \sum\limits_{j = 0}^\infty  {\frac{{( - i)^j }}{{j!}}} \int\limits_{ - \infty }^t {dt_j dt_{j - 1} ...dt_1 T_t \left\{ {A(t_1 )A(t_2 )...A(t_j )\rho _0 } \right\}} 
\end{eqnarray}
Substituting Eq.~(\ref{energy12}) into Eq.~(\ref{energy05})  we obtain the formula for  mean current
\begin{eqnarray}
\label{energy13}
 < I >  = \frac{1}{{Tr(\rho _I (t))}}\sum\limits_{j = 0}^\infty  {\frac{{( - i)^j }}{{j!}}} \int\limits_{ - \infty }^t {dt_j dt_{j - 1} ...dt_1 } Tr(T_t \left\{ {I(t)A(t_1 )A(t_2 )...A(t_j )\rho _0 } \right\})
\end{eqnarray}
This expression is essentially simplified by using the linked-diagram  theorem \cite{A11}, which excludes from consideration the  diagrams containing unlinked  blocks. This theorem is typical in quantum field methods of statistical physics. Let us  introduce the notation
\[
\frac{1}{{Tr(\rho _I (t))}}Tr(T_t \left\{ {I(t)A(t_1 )A(t_2 )...A(t_j )\rho _0 } \right\}) =  < T_t \left\{ {I(t)A(t_1 )A(t_2 )...A(t_j )\rho _0 } \right\} > _{0{\kern 1pt} {\kern 1pt} con.} 
\]
for the contributions from linked diagrams. Then Eq.~(\ref{energy13})  takes the  most simple form
\begin{eqnarray}
\label{energy14}
 < I >  = \sum\limits_{j = 0}^\infty  {( - i)^j } \int\limits_{ - \infty }^t {dt_j dt_{j - 1} ...dt_1 }  < T_t \left\{ {I(t)A(t_1 )A(t_2 )...A(t_j )\rho _0 } \right\} > _{0{\kern 1pt} {\kern 1pt} con.} 
\end{eqnarray}
Expression ~(\ref{energy14}) is the main when we consider an influence of the nonlinear contributions of perturbation theory on transport in the electron systems.

\section{Tunnel barrier in the hybrid structure :  normal metal -barrier- normal metal}

Let us consider  a simple task on electron transport in the system to be consisted from two  layers of normal metals which are  separated by oxide film creating the tunnel barrier. In Fig.1 the energy structure  of  a given model is figured and one can be presented in the form of Hamiltonian
\begin{eqnarray}
\label{energy15}
\hat H = \hat H_0  + H_T,
\end{eqnarray}
where $\hat H_0  = \hat H_L  + \hat H_R$. The unperturbed parts $\hat H_L  = \sum\limits_{\bm k} {(\varepsilon _{\bm k}  - \mu _L )n_{\bm k} }  = \sum\limits_{\bm p} {\xi _{\bm k} n_{\bm k} }$ and $\hat H_R  = \sum\limits_{\bm p} {(\varepsilon _{\bm p}  - \mu _R )n_{\bm p} }  = \sum\limits_{\bm p} {\xi _{\bm p} n_{\bm p} }$. Here, ${\varepsilon _{\bm k}}$ and ${\varepsilon _{\bm p}}$ are electron energies. The number operators $n_{\bm k}  = a_{\bm k}^ +  a_{\bm k}$ and $n_{\bm p}  = a_{\bm p}^ +  a_{\bm p}$ for electron states ${\bm k}$ and ${\bm p}$ and also the chemical potentials ${\mu_L}$ and ${\mu_R}$  are designated for left- and right-hand sides of hybrid structure, respectively.

The applied electrical voltage \textit{V} shifts the chemical potentials relatively one another in a such  manner that the relation  $\mu _L  - \mu _R  = eV$    is fulfilled, where \textit{e} is a modulus of the electron charge. The perturbation Hamiltonian ${H_T}$  describing the electron tunneling from left-  to right-hand side of hybrid structure has a form 
\begin{eqnarray}
\label{energy16}
H_T  = \sum\limits_{\bm {kp}} {\left( {T_{\bm {kp}} a_{\bm k}^ +  a_{\bm p}  + T_{\bm {kp}}^ *  a_{\bm p}^ +  a_{\bm k} } \right)}
\end{eqnarray}
\begin{figure}[h]
\begin{center}\resizebox{8.5cm}{!}{\includegraphics[width=7cm]{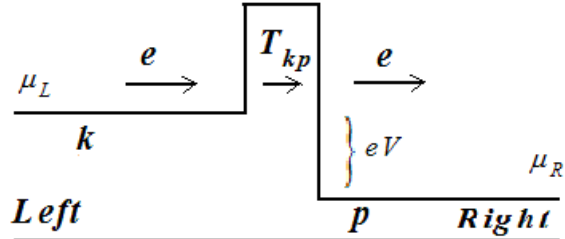}}
\end{center}
\caption{The energy structure of the model normal metal-barrier-normal metal.}
\label{Figure 1.}
\end{figure}

To evaluate a mean electrical current it is necessary to find the expression for operators ${A(t_j)}$  entering in Eq.~(\ref{energy14}). From the whole set of states with wave vectors ${\bm k}$ we will fix one ${\bm k}$. Then one can consider subspace of two wave functions $\left| {1_{\bm k} } \right\rangle$  and  $\left| {0_{\bm k} } \right\rangle$  with and without  electron in ${\bm k}$ -state, respectively. It is obvious that space of all states of the wave vectors is the direct product of these subspaces for each vector ${\bm k}$.  For example, the operator exponent for the left-hand side of the metal, which determines the unperturbed density matrix, can be represented as a direct product:
\begin{eqnarray}
\label{energy17}
e^{ - \beta \hat H_L }  = e^{ - \beta \sum\limits_{\bm k} {\xi _{\bm k} n_{\bm k} } }  = \prod\limits_{\bm k} {\left\{ {E_{\bm k}  + \left( {e^{ - \beta \varepsilon _{\bm k} }  - 1} \right)n_{\bm k} } \right\}}, 
\end{eqnarray}
where $E_{\bm k}  = \left( {\begin{array}{*{20}c}
   1 & 0  \\
   0 & 1  \\
\end{array}} \right)
$ is the unit  matrix and $n_{\bm k}  = \left( {\begin{array}{*{20}c}
   1 & 0  \\
   0 & 0  \\
\end{array}} \right)
$. Obviously, that
\begin{eqnarray}
\label{energy18}
Tr(e^{ - \beta \hat H_L } ) = \prod\limits_{\bm k} {\left\{ {1 + e^{ - \beta \varepsilon _{\bm k} } } \right\}}
\end{eqnarray}

One can simplify the expression (10) using Eq.~(\ref{energy06}), i.e. the task is reduced to calculation of the commutators of   ${\hat H_0}$ and  ${H_T}$. Indeed, it is not difficult to find a commutator in the first order of expansion ~(\ref{energy06}), which is expressed as follows:
\[
\left[ {\hat H_0 ,H_T } \right] = \sum\limits_{\bm {kp}}{(\xi _{\bm k}  - \xi _{\bm p} )\left\{ {T_{\bm {kp}}a_{\bm k}^ +  a_{\bm p}  - T_{\bm {kp}}^ *  a_{\bm p}^ +  a_{\bm k} } \right\}} 
\]
It can be shown that the commutator of the next order
\[
\left[ {\hat H_0 \left[ {\hat H_0 ,H_T } \right]} \right] = \sum\limits_{\bm {kp}}{(\xi _{\bm k}  - \xi _{\bm p} )^2 \left\{ {T_{\bm {kp}} a_{\bm k}^ +  a_{\bm p}  + T_{\bm {kp}}^ *  a_{\bm p}^ +  a_{\bm k} } \right\}} 
\]

The structure of the commutators of higher order is evident. Therefore, we immediately write the operator ${A(t_j)}$  from Eq.~(\ref{energy10}):
\begin{eqnarray}
\label{energy19}
\begin{array}{l}
 A(t_j ) =  \\ 
 \quad \quad \sum\limits_{\bm{kp}} {\left\{ {\left( {1 - e^{\beta (\xi _{\bm p}  - \xi _{\bm k} )} } \right)T_{\bm{kp}} a_{\bm k}^ +  (t_j )a_{\bm p} (t_j ) + \left( {1 - e^{ - \beta (\xi _{\bm p}  - \xi _{\bm k} )} } \right)T_{\bm{kp}}^ *  a_{\bm p}^ +  (t_j )a_{\bm k} (t_j )} \right\}}  \\ 
 \end{array}
\end{eqnarray}
This operator can not be substituted in Eq.~(\ref{energy14}) for calculation   $< I >$ , because in  the correlators it is necessary to carry out convolutions with respect to the indices ${\bm k}$ and ${\bm p}$. On the other hand, these same indices are also included in  ${\rho _0}$  of  expression ~(\ref{energy14}). Therefore, we select the factors with these indices from  ${\rho _0}$ and substitute ones in Eq.~(\ref{energy19}) that leads to a replacement ${A(t_j)}$ in Eq.~(\ref{energy14})   by ${\tilde A(t_j)}$ . This is easy to do if we take into account that the factors  in $\rho_0$ with the indices ${\bm k}$ and ${\bm p}$  have the form
\[
\frac{{\left( {E_{\bm k}  + \left( {e^{ - \beta \xi _{\bm k} }  - 1} \right)n_{\bm k} } \right)\left( {E_{\bm p}  + \left( {e^{ - \beta \xi _{\bm p} }  - 1} \right)n_{\bm p} } \right)}}{{Tr\left\{ {\left( {E_{\bm k}  + \left( {e^{ - \beta \xi _{\bm k} }  - 1} \right)n_{\bm k} } \right)\left( {E_{\bm p}  + \left( {e^{ - \beta \xi _{\bm p} }  - 1} \right)n_{\bm p} } \right)} \right\}}} = \kern 100pt\]  \  \[\frac{{\left( {E_{\bm k}  + \left( {e^{ - \beta \xi _{\bm k} }  - 1} \right)n_{\bm k} } \right)\left( {E_{\bm p}  + \left( {e^{ - \beta \xi _{\bm p} }  - 1} \right)n_{\bm p} } \right)}}{{\left( {1 + e^{ - \beta \xi _{\bm k} } } \right)\left( {1 + e^{ - \beta \xi _{\bm p} } } \right)}}
\]
As a result of this substitution  we obtain the expression for ${\tilde A(t_j)}$ :
\begin{eqnarray}
\label{energy20}
\tilde A(t_j ) = \sum\limits_{\bm {kp}} {\left[ {f(\xi _{\bm p} ) - f(\xi _{\bm k})} \right]\left\{ {T_{\bm {kp}} a_{\bm k}^ +  (t_j )a_{\bm p} (t_j ) - T_{\bm {kp}}^ *  a_{\bm p}^ +  (t_j )a_{\bm k} (t_j )} \right\}}, 
\end{eqnarray}
which one is substituted in Eq.~(\ref{energy14})  instead  of ${A(t_j)}$ . Now in the density matrix ${\rho _0}$  the factors with indices ${\bm k}$ and ${\bm p}$  are absent, i.e.  ${\rho _0}$ is replaced by ${\tilde\rho _0}$ . 

It should be noted that the theory presented in Section 3 described the evolution of the density matrix for a system with a fixed number of particles. In reality, there are two subsystems with chemical potentials  $\mu _L$ and  $\mu _R$. Due to the current, the number of particles at time \textit{t} in the left- and right-hand sides is not fixed. Therefore, evolution in a system with current must be described by the complete Hamiltonian \cite{A8}:
\begin{eqnarray}
\label{energy21}
\hat H' = \hat H'_0  + H_T , 
\end{eqnarray}
where
\[
\hat H'_0  = \sum\limits_{\bm k} {\varepsilon _{\bm k} n_{\bm k} }  + \sum\limits_{\bm p} {\varepsilon _{\bm p} n_{\bm p} }  = \sum\limits_{\bm k} {\xi _{\bm k} n_{\bm k} }  + \sum\limits_{\bm p} {\xi _{\bm p} n_{\bm p} }  + \mu _L N_L  + \mu _R N_R 
\]
${N_L  = \sum\limits_{\bm k} {n_{\bm k} }}$ and ${N_R  = \sum\limits_{\bm p} {n_{\bm p} } }$ is the total number of  electrons in the left-  and right-hand  side  of the given hybrid structure, respectively. Taking into account the above, the creation and annihilation operators in the expressions for ${H_T (t_j )}$  and  ${\tilde A(t_j)}$  take the form
\begin{eqnarray}
\label{energy22}
\begin{array}{l}
 \tilde a_{\bm k} (t_j ) = e^{ - it_j \mu _L } a_{\bm k} (t_j ) \\ 
 \tilde a_{\bm k}^ +  (t_j ) = e^{it_j \mu _L } a_{\bm k}^ +  (t_j ) \\ 
 \end{array}
 \end{eqnarray}
and  likewise for the right-hand side of the metal. Thus, we have
\begin{eqnarray}
\label{energy23}
H_T (t_j ) = \sum\limits_{\bm {kp}} {\left\{ {e^{ieVt_j } T_{\bm {kp}} a_{\bm k}^ +  (t_j )a_{\bm p} (t_j ) + e^{ - ieVt_j } T_{\bm {kp}}^ *  a_{\bm p}^ +  (t_j )a_{\bm k} (t_j )} \right\}} 
\end{eqnarray}
\begin{eqnarray}
\label{energy24}
\tilde A(t_j ) = \sum\limits_{\bm {kp}}{\left\{ {\tilde T_{1{\bm {kp}}} e^{ieVt_j } a_{\bm k}^ +  (t_j )a_{\bm p} (t_j ) + \tilde T_{2{\bm {kp}}} e^{ - ieVt_j } a_{\bm p}^ +  (t_j )a_{\bm k} (t_j )} \right\}}, 
\end{eqnarray}
where
\begin{eqnarray}
\label{energy25}
\begin{array}{l}
 \tilde T_{1{\bm {kp}}}  = \left[ {f(\xi _{\bm p} ) - f(\xi _{\bm k} )} \right]T_{\bm {kp}}  \\ 
 \tilde T_{2{\bm {kp}}}  =  - \tilde T_{1{\bm {kp}}}^ *   \\ 
 \end{array}
\end{eqnarray}
Since ${\frac{{d(N_L  + N_R )}}{{dt}} = 0}$ and ${\frac{{dN_L (t)}}{{dt}} = i\left[ {H,N_L } \right] =  - i\left[ {N_L ,H_T } \right]}$, then the  current operator is determined as 
\[
I(t) =  - e\frac{{dN_L (t)}}{{dt}}
\]
and
\begin{eqnarray}
\label{energy26}
I(t) = ie\sum\limits_{\bm {kp}} {\left\{ {e^{ieVt} T_{\bm {kp}} a_{\bm k}^ +  (t)a_{\bm p} (t) - e^{ - ieVt} T_{\bm {kp}}^ *  a_{\bm p}^ +  (t)a_{\bm k} (t)} \right\}} 
\end{eqnarray}
The initial density matrix ${\rho _0}$  is determined by the unperturbed  Hamiltonian $\hat H_0$  since  in each subsystems in absence current the particle numbers are fixed.

\section{Linear response in the metal with a tunnel barrier}

Here, we will consider the use of  Eq.~(\ref{energy14}) to find electrical current in a linear approximation. The expression ~(\ref{energy14}) takes the form
\begin{eqnarray}
\label{energy27}
 < I > _1  =  - i\int\limits_{ - \infty }^t {dt_1 }  < T_t \left\{ {I(t)\tilde A(t_1 )\tilde \rho _0 } \right\} > _{0{\kern 1pt} {\kern 1pt} con.} 
\end{eqnarray}
Substituting in Eq.~(\ref{energy27})  the expressions ~(\ref{energy24}) and ~(\ref{energy26}) we obtain the linear current
\begin{eqnarray}
\label{energy28}
\begin{array}{l}
  < I > _1  = \\ 2e\sum\limits_{\bm {kpk_1 p_1}} {\int\limits_{ - \infty }^t {dt_1 }  < T_t \left\{ {\tilde T_{1\bm{k_1 p_1}} e^{ieVt_1 } a_{{\bm k}_1 }^ +  (t_1 )a_{{\bm p}_1 } (t_1 ) + \tilde T_{2\bm{k_1 p_1}} e^{ - ieVt_1 } a_{{\bm p}_1 }^ +  (t_1 )a_{{\bm k}_1 } (t_1 )}  \right\}} \\  \quad \quad \times 
 \left\{ {e^{ieVt} T_{\bm {kp}} a_{\bm k}^ +  (t)a_{\bm p} (t) - e^{ - ieVt} T_{\bm {kp}}^ * a_{\bm p}^ +  (t)a_{\bm k} (t)} \right\}\tilde \rho _0  > _{0{\kern 1pt} {\kern 1pt} con.}  \\ 
\end{array}
\end{eqnarray}
Factor 2 arises from summing up over  electron spins. Keeping in Eq.~(\ref{energy28}) nonzero correlators only and taking into account for definitions ~(\ref{energy25}) we have
\begin{eqnarray}
\label{energy29}
\begin{array}{l}
  < I > _1  =  - 2e\sum\limits_{\bm {kpk_1 p_1}} {\left[ {f(\xi _{{\bm p}_1 } ) - f(\xi _{{\bm k}_1 } )} \right]}  \times  \\ 
 \quad \quad \int\limits_{ - \infty }^t {dt_1 } \left\{ {e^{i(t - t_1 )eV} T_{\bm {kp}} T_{\bm{k_1 p_1} }^ *   < T_t a_{\bm k}^ +  (t)a_{\bm p} (t)a_{{\bm p}_1 }^ +  (t_1 )a_{{\bm k}_1 } (t_1 )\tilde \rho _0  > _{0{\kern 1pt} {\kern 1pt} con.} } \right. +  \\ 
 \left. {\quad \quad \quad \quad T_{\bm{kp}}^ *  T_{\bm{k_1 p_1}} e^{ - i(t - t_1 )eV}  < Ta_{\bm p}^ +  (t)a_{\bm k} (t)a_{{\bm k}_1 }^ +  (t_1 )a_{{\bm p}_1 } (t_1 )\tilde \rho _0  > _{0{\kern 1pt} {\kern 1pt} con.} } \right\} \\ 
 \end{array}
\end{eqnarray}

The correlators obtained in Eq.~(\ref{energy29})  are decoupled by Wick's theorem which allows us to write down
\begin{eqnarray}
\label{energy30}
\begin{array}{l}
  < I > _1  =  - 2e\sum\limits_{\bm{kp}} {\left[ {f(\xi _{\bm p} ) - f(\xi _{\bm k} )} \right]} \left| {T_{\bm{kp}} } \right|^2  \times  \\ 
 \quad \quad \int\limits_{ - \infty }^t {dt_1 } \left\{ {e^{i(t - t_1 )eV}  < T_t a_{\bm k}^ +  (t)a_{\bm k} (t_1 ) > _0  < T_t a_{\bm p} (t)a_{\bm p}^ +  (t_1 ) > _{0{\kern 1pt} {\kern 1pt} } } \right. +  \\ 
 \left. {\quad \quad \quad \quad e^{ - i(t - t_1 )eV}  < T_t a_{\bm p}^ +  (t)a_{\bm p} (t_1 ) >  < T_t a_{\bm k} (t)a_{\bm k}^ +  (t_1 ) > _{0{\kern 1pt} } } \right\} \\ 
 \end{array}
\end{eqnarray}

In Eq.~(\ref{energy30}) the operator  ${\tilde\rho _0}$  disappears  since it has not multipliers  with indices ${\bm k}$ and ${\bm p}$. Also, in Eq.~(\ref{energy30})  there are unperturbed Green's functions  which one can  define as follows:
\begin{eqnarray}
\label{energy31}
\begin{array}{l}
 G_{\bm k} (t - t_1 ) =  -  < T_t a_{\bm k} (t)a_{\bm k}^ +  (t_1 ) > _{0{\kern 1pt} }  =  \\ 
 \quad  - \theta (t - t_1 ) < a_{\bm k} (t)a_{\bm k}^ +  (t_1 ) > _0  + \theta (t_1  - t) < a_{\bm k}^ +  (t_1 )a_{\bm k} (t) > _0  =  \\ 
 \quad \quad e^{i\xi _{\bm k} (t_1  - t)} \left( { - \theta (t - t_1 )Tr(E_{\bm k}  - n_{\bm k} ) + \theta (t_1  - t)Tr(n_{\bm k} )} \right) =  \\ 
 \quad \quad \quad e^{i\xi _{\bm k} (t_1  - t)} \left( { - \theta (t - t_1 ) + \theta (t_1  - t)} \right), \\ 
 \end{array}
\end{eqnarray}
since ${Tr(E_{\bm k}  - n_{\bm k} ) = Tr(n_{\bm k} ) = 1}$. The Fourier transform for ${G_{\bm k} (t - t_1 )}$   has a form:
\begin{eqnarray}
\label{energy32}
G_{\bm k} (\omega ) = \int\limits_{ - \infty }^{ + \infty } {G_{\bm k} (t - t_1 )e^{i\omega t} dt = \frac{1}{i}\left( {\frac{1}{{\omega  - \xi _{\bm k}  - i\delta }} + \frac{1}{{\omega  - \xi _{\bm k}  + i\delta }}} \right)}, 
\end{eqnarray}
where the infinitesimal imaginary corrections  are due to adiabatic switching on and off of the perturbation. The  Fourier transform  of the Green's functions in Eq.~(\ref{energy30})  gives
\[
\begin{array}{l}
  < I > _1  = 2e\sum\limits_{\bm{kp}} {\left[ {f(\xi _{\bm p} ) - f(\xi _{\bm k} )} \right]} \left| {T_{\bm{kp}} } \right|^2  \times  \\ 
 \quad \quad \;\;\int\limits_{ - \infty }^{ + \infty } {\frac{{d\omega _1 d\omega _2 }}{{(2\pi )^2 }}G_{\bm k} (\omega _1 )G_{\bm p} (\omega _2 )} \left\{ {\int\limits_{ - \infty }^t {dt_1 } \left( {e^{i(t_1  - t)(\omega _2  - \omega _1  - eV)}  + e^{ - i(t_1  - t)(\omega _2  - \omega _1  - eV)} } \right)} \right\} \\ 
 \end{array}
\]
As a result of integration over variable $t_1$ we obtain
\begin{eqnarray}
\label{energy33}
\begin{array}{l}
  < I > _1  = 2e\sum\limits_{\bm{kp}} {\left[ {f(\xi _{\bm p} ) - f(\xi _{\bm k} )} \right]\left| {T_{\bm{kp}} } \right|^2 \int\limits_{ - \infty }^{ + \infty } {\frac{{d\omega _1 d\omega _2 }}{{(2\pi )^2 }}G_{\bm k} (\omega _1 )G_{\bm p} (\omega _2 ) \times } }  \\ 
 \;\quad \quad \quad \quad \quad \quad \quad \left\{ {\frac{1}{{i(\omega _2  - \omega _1  - eV - i\delta )}} - \frac{1}{{i(\omega _2  - \omega _1  - eV + i\delta )}}} \right\} \\ 
 \end{array}
\end{eqnarray}

After substituting the  expressions for Fourier transforms of the Green functions in Eq.~(\ref{energy33}), it is easy to integrate with respect to the variables $\omega_1$ and $\omega_2$.  The contours of integration are closed  in the upper or lower half-planes of complex variables, depending on the presence of the minimal number of poles. Thus, we have
\begin{eqnarray}
\label{energy34}
\begin{array}{l}
  < I > _1  =  \\ 
 \quad 2e\sum\limits_{\bm{kp}} {\left[ {f(\xi _{\bm p} ) - f(\xi _{\bm k} )} \right]\left| {T_{\bm{kp}} } \right|^2 } \left\{ {\frac{1}{{i(\xi _{\bm p}  - \xi _{\bm k}  - eV + 3i\delta )}} - \frac{1}{{i(\xi _{\bm p}  - \xi _{\bm k}  - eV - 3i\delta )}}} \right\} \\ 
 \end{array}
\end{eqnarray}

In order to perform the summation in Eq.~(\ref{energy34}) over the wave vectors  let us assume $\left| {T_{kp} } \right|^2  = \left| T \right|^2$  and put  for simplicity the constant electron densities of the states $D_L$  and $D_R$  and denote  the integration variables $\xi_L$  and $\xi_R$  for the left-hand  and right-hand sides of the metal, respectively. In accordance with  the Landau rule for bypass the poles  one can write  expression  (34)  in  the form:
\begin{eqnarray}
\label{energy35}
\begin{array}{l}
  < I > _1  = 4\pi e\left| T \right|^2 D_L D_R \int\limits_{ - E_F }^{ + \infty } {d\xi _R \int\limits_{ - E_F }^{ + \infty } {d\xi _L \left[ {f(\xi _L ) - f(\xi _R )} \right]\delta \left( {\xi _L  - \xi _R  + eV} \right)} }  =  \\ 
 \quad \quad \quad \quad \quad \quad \quad 4\pi e\left| T \right|^2 D_L D_R \int\limits_{ - E_F }^{ + \infty } {d\xi _R \left[ {f(\xi _R  - eV) - f(\xi _R )} \right]},  \\ 
 \end{array}
\end{eqnarray}
where $\delta(x)$  is  the Dirac delta function. Let us consider the case of  temperatures $\quad \quad$ $T << E_F$  where  $E_F$ is the Fermi energy. Then the Fermi distribution function can be replaced by the Heaviside step function $\theta(-x)$. As a result, we have
\begin{eqnarray}
\label{energy36}
\begin{array}{l}
  < I > _1  =  \\
 4\pi e\left| T \right|^2 D_L D_R \int\limits_{ - E_F }^{ + \infty } {d\xi _R \left[ {\theta ( - (\xi _R  - eV)) - \theta ( - \xi _R )} \right]}  =  4\pi e\left| T \right|^2 D_L D_R \int\limits_0^{eV} {d\xi _R }  = \\ \quad \quad\quad \quad\quad \quad\quad \quad \kern 200pt 4\pi e^2 \left| T \right|^2 D_L D_R V \\  
 \end{array}
\end{eqnarray}
This expression reflects the Ohm's law and coincides with a  similar  formula presented in \cite{A8} within the framework of the linear Kubo theory for electron transport.

\section{Nonlinear contribution in the electron transport}

Using Eq.~(\ref{energy14})  it is not difficult to generalize an expression for the current in any order of perturbation theory. Obviously, the next nonzero contribution appears only in the third order since the second order of series expansion ~(\ref{energy14}) includes an odd number of  creation  and annihilation operators for each metal subsystems. One can write the contribution of the third order from expression ~(\ref{energy14}):
\begin{eqnarray}
\label{energy37}
 < I > _1  =  - i\int\limits_{ - \infty }^t {dt_1 }{dt_2 }{dt_3 }  < T_t \left\{ {I(t)\tilde A(t_1 )\tilde A(t_2 )\tilde A(t_3 )\tilde \rho _0 } \right\} > _{0{\kern 1pt} {\kern 1pt} con.} 
\end{eqnarray}
Substituting in Eq.~(\ref{energy37})  the expression for ${\tilde A(t_j)}$  from Eq.~(\ref{energy19}) we have
\begin{eqnarray}
\label{energy38}
\begin{array}{l}
  < I > _3  =  - 4e\sum\limits_{\bm{^{kpk_1 p_1}_{k_2 p_2 k_3 p_3}}} {\int\limits_{ - \infty }^t {dt_1 dt_2 dt_3 } } \left\{ {e^{i(t - t_1  + t_2  - t_3 )eV} T_{\bm{kp}} \tilde T_{\bm{2k_1 p_1 }} \tilde T_{\bm{1k_2 p_2 }}^{} \tilde T_{\bm{2k_3 p_3} }  \times } \right. \\ 
 \quad  < T_t a_{\bm k}^ +  (t)a_{\bm p} (t)a_{{\bm p}_1 }^ +  (t_1 )a_{\bm{k_1} } (t_1 )a_{{\bm k}_2 }^ +  (t_2 )a_{{\bm p}_2 } (t_2 )a_{{\bm p}_3 }^ +  (t_3 )a_{{\bm k}_3 } (t_3 )\tilde \rho _0  > _{0{\kern 1pt} {\kern 1pt} con.}  -  \\ 
 \quad \quad \quad \quad \quad e^{ - i(t - t_1  + t_2  - t_3 )eV} T_{\bm{kp}}^ *  \tilde T_{\bm{1k_1 p_1} } \tilde T_{\bm{2k_2 p_2} } \tilde T_{\bm{1k_3 p_3} }  \times  \\ 
 \quad \left. { < Ta_{\bm p}^ +  (t)a_{\bm k} (t)a_{{\bm k}_1 }^ +  (t_1 )a_{{\bm p}_1 } (t_1 )a_{{\bm p}_2 }^ +  (t_2 )a_{{\bm k}_2 } (t_2 )a_{{\bm k}_3 }^ +  (t_3 )a_{{\bm p}_3 } (t_3 )\tilde \rho _0  > _{0{\kern 1pt} {\kern 1pt} con.} } \right\} \\ 
 \end{array}
\end{eqnarray}
An additional factor of 4  arises from summing up over spin  indices. Carrying out the decoupling of the correlators entering  in Eq.~(\ref{energy38}),   we obtain
\begin{eqnarray}
\label{energy39}
\begin{array}{l}
  < I > _3  = 4e\sum\limits_{\bm {kpk_1 p_1}} {\varphi _{\bm {kpk_1 p_1}}} \int\limits_{ - \infty }^t {dt_1 dt_2 dt_3  \times }  \\ 
 \quad \quad \quad \quad \left\{ {e^{i(t - t_1  + t_2  - t_3 )eV} G_{\bm k} (t_3  - t)G_{\bm p} (t - t_1 )G_{{\bm k}_1 } (t_1  - t_2 )G_{{\bm p}_1 } (t_2  - t_3 ) + } \right. \\ 
 \left. {\quad \quad \quad \quad e^{ - i(t - t_1  + t_2  - t_3 )eV} G_{\bm k} (t - t_3 )G_{\bm p} (t_1  - t)G_{{\bm k}_1 } (t_2  - t_1 )G_{{\bm p}_1 } (t_3  - t_2 )} \right\}, \\ 
 \end{array}
\end{eqnarray}
where $\varphi _{\bm{kpk_1 p_1 }}  = \left| T \right|^4 \left[ {f(\xi _{\bm p} ) - f(\xi _{{\bm k}_1 } )} \right]\left[ {f(\xi _{{\bm p}_1 } ) - f(\xi _{{\bm k}_1 } )} \right]\left[ {f(\xi _{{\bm p}_1 } ) - f(\xi _{\bm k} )} \right]$. As in $\quad$ Section 3,  the Fourier transform of  the Green functions  and  the following  integration over $t_1$, $t_2$  and  $t_3$  allows to write
\begin{eqnarray}
\label{energy40}
\begin{array}{l}
  < I > _3  = 4e\sum\limits_{\bm{kpk_1 p_1} } {\varphi _{\bm{kpk_1 p_1 }} } \int\limits_{ - \infty }^t {d\omega _1 d\omega _2 d\omega _3 d\omega _4 } \left\{ {G_{\bm k} (\omega _1 )G_{\bm p} (\omega _2 )G_{{\bm k}_1 } (\omega _3 )G_{{\bm p}_1 } (\omega _4 ) \times } \right. \\ 
 \quad \quad \quad \quad \quad \quad \left. {\left\{ {\Psi (\omega _1 ,\omega _2 ,\omega _3 ,\omega _4 ,  \delta ) - \Psi (\omega _1 ,\omega _2 ,\omega _3 ,\omega _4 , - \delta )} \right\}} \right\}, \\ 
 \end{array}
\end{eqnarray}
where 
\[
\Psi (\omega _1 ,\omega _2 ,\omega _3 ,\omega _4 ,\delta ) = \frac{1}{{i(\omega _2  - \omega _3  - eV - i\delta )}}\frac{1}{{i(\omega _3  - \omega _4  + eV - i\delta )}}\frac{1}{{i(\omega _4  - \omega _1  - eV - i\delta )}}
\]

From Eq.~(\ref{energy40}) it follows that $<I>_3$   as $<I>_1$  are determined by difference of the complex conjugate values,  i.e. $<I>_3$  is real and equal to double  imagine part of  one of the integrals. Obviously, this is due to the structure of the current operator, which describes the electron hopping through the barrier in one as well as in the opposite directions independently on the order of the perturbation theory. The integrals in Eq.~(\ref{energy40}) are calculated using the theory of residues. Denoting the fourfold integral in Eq.~(\ref{energy40}) through $J_{\bm{k,p,k_1,p_1}}(\delta)$ , we can write its expression in the form:
\begin{eqnarray}
\label{energy41}
J_{kpk_1 p_1 } (\delta ) = \sum\limits_{i = 1}^4 {(J_{kpk_1 p_1 }^i (\delta ) - J_{kpk_1 p_1 }^i ( - \delta ))} , 
\end{eqnarray}
where
\begin{eqnarray}
\label{energy42}
\begin{array}{l}
 J_{\bm{kpk_1 p_1} }^1 (\delta ) =  - \frac{1}{{i(\xi _{\bm p}  - \xi _{{\bm k}_1 }  - eV - 3i\delta )}}\frac{1}{{i(\xi _{\bm p}  - \xi _{{\bm p}_1 }  - 4i\delta )}}\frac{1}{{i(\xi _{\bm p}  - \xi _{\bm k}  - eV - 5i\delta )}} \\ 
 J_{\bm{kpk_1 p_1} }^2 (\delta ) =  - \frac{1}{{i(\xi _{\bm p}  - \xi _{{\bm k}_1 }  - eV - 3i\delta )}}\frac{1}{{i(\xi _{\bm k}  - \xi _{{\bm p}_1 }  + eV + 3i\delta )}}\frac{1}{{i(\xi _{\bm p}  - \xi _{\bm k}  - eV - 5i\delta )}} \\ 
 J_{\bm{kpk_1 p_1} }^3 (\delta ) = \frac{1}{{i(\xi _{{\bm p}_1 }  - \xi _{{\bm k}_1 }  - eV + 3i\delta )}}\frac{1}{{i(\xi _{{\bm p}_1 }  - \xi _{\bm p}  + 4i\delta )}}\frac{1}{{i(\xi _{{\bm p}_1 }  - \xi _{\bm k}  - eV - i\delta )}} \\ 
 J_{\bm{kpk_1 p_1} }^4 (\delta ) = \frac{2}{{i(\xi _{\bm k}  - \xi _{{\bm k}_1 }  + 4i\delta )}}\frac{1}{{i(\xi _{\bm k}  - \xi _{{\bm p}_1 }  + eV + i\delta )}}\frac{1}{{i(\xi _{\bm k}  - \xi _{\bm p}  + eV + 5i\delta )}} \\ 
 \end{array}
\end{eqnarray}

The factor 2 in the last term of Eq.~(\ref{energy42}) appears due to the fact that in Eq.~(\ref{energy41}) there should be 5 terms, but 2 of them are the same. It is seen from Eq.~(\ref{energy42}) that the poles are located both in the upper and lower complex planes. For calculation of the current in a limit  $\delta\rightarrow 0$  the numerical factors in front of $\delta$ does not play any role. Also, in Eq.~(\ref{energy41}) only the differences of complex conjugate quantities enter. Thus, in the expression for $<I>_3$  the integrals in the sense of a principal value are absent. By analogy with Section 3, taking into account the Landau rule for bypassing poles we have
\[
\begin{array}{l}
  < I > _3  = 8\pi ^3 e\sum\limits_{\bm{kpk_1 p_1} } {\left| T \right|^4 \left[ {f(\xi _{\bm p} ) - f(\xi _{{\bm k}_1 } )} \right]\left[ {f(\xi _{{\bm p}_1 } ) - f(\xi _{{\bm k}_1 } )} \right]\left[ {f(\xi _{{\bm p}_1 } ) - f(\xi _{\bm k} )} \right]}  \times  \\ 
 \left\{ { - \delta (\xi _{{\bm p}_1 }  - \xi _{\bm p} )\delta (\xi _{{\bm k}_1 }  - \xi _{\bm p}  + eV)\delta (\xi _{\bm p}  - \xi _{\bm k}  - eV) + } \right. \\ 
 \quad \delta (\xi _{{\bm k}_1 }  - \xi _{\bm p}  + eV)\delta (\xi _{{\bm p}_1 }  - \xi _{\bm k}  - eV)\delta (\xi _{\bm p}  - \xi _{\bm k}  - eV) +  \\ 
 \quad \quad \delta (\xi _{{\bm p}_1 }  - \xi _{\bm p} )\delta (\xi _{\bm k}  - \xi _{{\bm p}_1 }  + eV)\delta (\xi _{{\bm p}_1 }  - \xi _{{\bm k}_1 }  - eV) -  \\ 
 \quad \quad \quad \left. {2\delta (\xi _{{\bm k}_1 }  - \xi _{\bm k} )\delta (\xi _{{\bm p}_1 }  - \xi _{\bm k}  - eV)\delta (\xi _{\bm p}  - \xi _{\bm k}  - eV)} \right\} \\ 
 \end{array}
\]

Assuming that the electronic densities of states are constant, we replace the  sums over wave vectors by  integrals  with the same integration variables as in Section 3. It is easy to find that
\begin{eqnarray}
\label{energy43}
\begin{array}{l}
  < I > _3  = 8\pi ^3 e\left| T \right|^4 D_L^2 D_R^2 \int\limits_{ - E_F }^{ + \infty } {d\xi _R \left[ {f(\xi _R  - eV) - f(\xi _R )} \right]^3 }  =  \\ 
 \quad \quad \quad \quad \quad \quad \quad \quad 8\pi ^3 e^2 \left| T \right|^4 D_L^2 D_R^2 \int\limits_0^{eV} {d\xi _R }  = 8\pi ^3 e^2 \left| T \right|^4 D_L^2 D_R^2 V \\ 
 \end{array}
\end{eqnarray}

Apparently, in the fifth  order of the perturbation theory we have the contribution in a mean electrical current
\[
 < I > _5  = 16\pi ^5 e\left| T \right|^6 D_L^3 D_R^3 \int\limits_{ - E_F }^{ + \infty } {d\xi _R \left[ {f(\xi _R  - eV) - f(\xi _R )} \right]^5 } 
\]
and so on. Summing up the contribution for all orders we obtain  the geometric series at condition $2\pi ^2 \left| T \right|^2 D_L D_R \left[ {f(\xi _L ) - f(\xi _R )} \right]^2  < 1$. Thus, the common factor of all the contributions to the current is twice the geometric ratio. Then the general expression for the mean current $<I>$  at finite temperatures takes the form:
\begin{eqnarray}
\label{energy44}
 < I >  = 4\pi e\left| T \right|^2 D_L D_R \int\limits_{ - E_F }^{ + \infty } {d\xi _R \frac{{f(\xi _R  - eV) - f(\xi _R )}}{{1 - 2\pi ^2 \left| T \right|^2 D_L D_R^{} \left[ {f(\xi _R  - eV) - f(\xi _R )} \right]^2 }}} 
\end{eqnarray}
In particular, at zero temperature a mean electrical current is presented in form of the renormalized Landauer \cite{A12} formula with $\hbar\neq 1$ :
\begin{eqnarray}
\label{energy45}
 < I >  = \frac{{4\pi e^2 }}{\hbar }\frac{{\left| T \right|^2 D_L D_R }}{{1 - 2\pi ^2 \left| T \right|^2 D_L D_R }}V = \frac{{2e^2 }}{h}\frac{{2\tilde T}}{{1 - \tilde T}}V,
\end{eqnarray}
from which it follows that barrier  transmission  must be determined as $\tilde T = 2\pi ^2 \left| T \right|^2 D_L D_R$. It is connected with the  probability of the barrier tunneling and depends on electron densities of the  left- and right-hand sides of   metal.
 
From  Eq.~(\ref{energy45})   it follows that for barrier transmission $\tilde T = 1$     the current becomes  infinity since the scattering centers are absent. In a single electron approximation the presence of the finite Landauer resistance $R_k  = \frac{h}{{2e^2 }}$  at full barrier transparency gives rise to contradiction that are explained by many- channel processes in the leads \cite{A5,A6}. However, from general Eq.~(\ref{energy45}) it follows  that with  the nonlinear contributions the pointed contradiction is removed. The  barrier transmission is determined by repeated electron penetrations and reflections. The electron statistic does not allow to  separate the penetration and reflection electron processes. Thus, one can say about effective barrier transmission in hybrid structures. Only at $\tilde T = 1/3$    we obtain the ballistic transport with Landauer conductivity $G = \frac{{2e^2 }}{h}$. Also, the account for electron quantum statistics leads to the fact   that the effective coefficient of electron reflection ${\tilde R}$  from the barrier can not be represented in a form  $\tilde R = 1 - \tilde T$  as for  one-electron approximation \cite{A13}.

It can be seen from Eq.~(\ref{energy45}) that the value $\left| T \right|$  of the tunnel matrix element that determines the probability of penetration of an electron through a barrier can not exceed a value $\frac{1}{{2\pi \sqrt {D_L D_R } }}$  . For this maximum value  the transport of electrons is equivalent to transport in systems without a barrier and scattering. The form of  Eq.~(\ref{energy44}), apparently, removes the contradiction of use  the tunnel Hamiltonian \cite{A8}, where it was indicated that the exponential growth $\left| T \right|^2$  as a function $\xi_{\bm k}$  can exceed the power-law drop of the spectral density. Then the main contribution to transport is caused by the top of the barrier. That is an unphysical result. As seen from Eq.~(\ref{energy44}), $\left| T \right|^2$   enters both in the numerator and the denominator. It  significantly complicates the relationship between the spectral density and  height of the barrier.

\section{Summary}

The paper presents a modified nonlinear Kubo theory for electron transport in hybrid structures of normal metals with a tunnel barrier. The results of the work can be generalized to the case of superconducting layers and Josephson junctions. The general expression for a mean current is obtained with account  for all contributions of the time-dependent perturbation theory. It is found that the nonlinearity of the theory causes a substantial renormalization of the Landauer formula for conductivity in the ground state. It turns out to be the Landauer formula works  in the framework of the one-electron approximation only. The ballistic transport conductivity $G = \frac{{2e^2 }}{h}$  is realized for barrier transmission $\tilde T = 1/3$.  The validity of the tunneling Hamiltonian for transport phenomena in hybrid structures is also justified.

\section*{Acknowledgements}

It is a pleasure to acknowledge a number of stimulating discussions with E.M. Rudenko, M.A.Belogolovskii  and A. P. Shapovalov.

\section*{Funding}

The study was carried out within the Fundamental Research Programme funded by the MES of Ukraine (Project No. 0117U002360).

\end{document}